\documentstyle[aps]{revtex}
\begin{document}
\centerline{\Large On the Matrix Element of the Transverse Component 
of}
\centerline{\Large Bilocal 
Vector Current and its Parton Interpretation}
\vskip .1in
\centerline {\large A. Harindranath$^{a}$, 
Wei-Min Zhang$^{b}$}
\vskip .1in
\centerline {\it $^{a}$Saha Institute of Nuclear Physics,
1/AF Bidhan Nagar, 
Calcutta, 700064 India} 
\centerline{\it $^{b}$Department of Physics, National Tsing Hua University,
Hsinchu, 30043 Taiwan}
\date{June 21, 1996}

\begin{abstract}
In this paper we study the matrix element of the transverse component of the
bilocal vector current in the context of deep inelastic scattering. 
BJL limit of high energy amplitudes together with light-front current
algebra imply the same parton interpretation for its matrix element as that
of the plus component. On the other hand, the transverse component depends
explicitly on the gluon field operator in QCD, appears as $``$twist three"
and hence its matrix element has no manifest parton interpretation. In this 
paper we perform calculations in light-front time-ordering perturbative QCD 
for a dressed quark target to order $\alpha_s$ and demonstrate that the 
matrix element of the transverse component of the bilocal vector current 
has the same parton interpretation as that of the plus component.
\end{abstract}
\centerline{PACS: 11.10Ef; 11.40.-q; 12.38.Bx}
{\it Keywords: bilocal vector current; parton picture;
light-front time-ordering perturbative QCD}

\vspace{.5in}
The bilocal vector current originated in the context of BJL limit and 
light-front current algebra \cite{gf,jackiw,dffr}. It plays a significant
role in the current understanding of factorization \cite{css} and power
corrections \cite{jj,jaffe} in deep inelastic scattering.
The hadronic matrix element of the bilocal vector current can be reduced to 
two bilocal form factors ${\bar V}_1$ and ${\bar V}_2$ defined by
\begin{eqnarray}
\langle P \mid {\bar {\cal V}}^\mu (y \mid 0) \mid P \rangle = 
P^\mu {\bar V}_1(y^2, P.y) + y^\mu {\bar V}_2(y^2, P.y), \label{def0}
\end{eqnarray}
where 
\begin{eqnarray}
{\bar {\cal V}^\mu }(y \mid z) = { 1 \over 2 i} \big 
[ V^\mu (y \mid z) - V^\mu (z\mid y) \big],
\end{eqnarray}
with the bilocal vector current 
\begin{eqnarray}
V^\mu(y \mid z) = {\bar \psi}(y) \gamma^\mu \psi(z),
\end{eqnarray}
$\psi$ being the fermion field.

The bilocal form factors relevant for deep inelastic scattering are
associated with $y^+=0$ and $y^\perp=0$.
Thus $y^2=0$ and $P \cdot y = { 1 \over 2} P^+y^-$.  
The twist two part of the unpolarized deep inelastic structure function
$F_2$ is defined by 
\begin{eqnarray}
{F_2(x) \over x} = { i \over 4 \pi}P^+ \int dy^- e^{-{ i \over 2}P^+ y^-x}
{\bar V}_1(y^-).  \label{f2} 
\end{eqnarray} 

Usually, the nucleon target is considered in the rest frame, $P^\mu
=(M, 0)$, so that $P^\bot =0$. Since $y^+=0$, ${\bar V}_1$ can be 
extracted using the plus component of the bilocal current 
${\bar {\cal V}^+}$ from Eq. (\ref{def0}), which leads to
\begin{eqnarray}
{F_2(x)  \over x} = { 1 \over 8 \pi} \int dy^- e^{-{i \over 2} P^+ y^- x} 
\langle P \mid \big [ {\bar \psi}(y) \gamma^+ \psi(0) - {\bar \psi}(0)
\gamma^+  \psi(y) \big ] |P\rangle.  \label{def1}
\end{eqnarray}
This is the familiar expression of $F_2$ in terms of hadronic matrix 
element of the plus component of light-front bilocal vector current.
In this case the operator has no explicit dependence on the interaction 
(the full complexities of the interaction are hidden in the state)
and the parton interpretation of the $F_2$ structure function is manifest
using the light-front formalism\cite{soper,css}.

However, Eqs.(\ref{def0}) and (\ref{f2}) are also generally true for 
the nucleon target being in any arbitrary Lorentz frame. If we consider
an arbitrary frame in which $P^\bot \neq 0$, then it follows from 
Eq. (\ref{def0}) that ${\bar V}_1$ can also be extracted from   
the transverse component ${\bar {\cal V}^\perp}$ since $y^\perp=0$
and Eq.(\ref{f2}) can be expressed by 
\begin{eqnarray}
{F_2(x) \over x} = { 1 \over 8 \pi} {P^+ \over P^\perp} \int dy^- 
e^{- {i \over 2} P^+y^-x} \langle P \mid {\bar \psi}(y) \gamma^\perp 
\psi(0) - {\bar \psi}(0) \gamma^\perp \psi(y) \mid P \rangle. \label{def2}
\end{eqnarray}
Now the operator that appears in Eq. (\ref{def2})
\begin{eqnarray}
{\bar \psi}(y) \gamma^\perp \psi(0) 
= (\psi^+)^\dagger(y) \alpha^\perp \psi^-(0) 
+ (\psi^-)^\dagger(y) \alpha^\perp \psi^+(0) \label{perp}
\end{eqnarray}
is explicitly interaction dependent because of the presence of the
constrained field operator $\psi^-$.  
Thus the operator
${\bar {\cal V}^\perp}$ has explicit dependence on the interaction and the
structure function $F_2$ extracted from Eq. (\ref{def2}) has no manifest
parton interpretation.
In fact, due to the explicit dependence on the gluon field,
the operator ${\bar {\cal V}^\perp}$ appears to be of $``$higher twist"
\cite{jj,jaffe}.

To the best of our knowledge, this is for the first time  
it is observed that the
unpolarized structure function $F_2$ can also be expressed in terms of
the transverse component of the light-front bilocal vector current,
Eq.(\ref{def2}). 
Since Eqs.(\ref{def1}) and (\ref{def2}) describe the same structure 
function, it is necessary to check their 
internal consistency. In this letter, we demonstrate the internal 
consistency and hence the parton interpretation of the transverse 
component of the bilocal vector current by explicitly carrying out
calculations of the structure function of a dressed quark up to order
$\alpha_s$. Note that for a quark target, contribution from the 
second term in Eqs. (\ref{def1}) and (\ref{def2}) vanishes.  

In order to evaluate Eqs. (\ref{def1}) and (\ref{def2}) for a 
quark target to order $\alpha_s$, we take the state 
$ \mid P \rangle$ to be a dressed quark
and expand this state in terms of bare states of quark and quark plus 
gluon:
\begin{eqnarray}
\mid P, \sigma \rangle && = \phi_1 b^\dagger(p,\sigma) \mid 0 \rangle
\nonumber \\  
&& + \sum_{\sigma_1,\lambda_2} \int 
{dk_1^+ d^2k_1^\perp \over \sqrt{2 (2 \pi)^3 k_1^+}}  
\int 
{dk_2^+ d^2k_2^\perp \over \sqrt{2 (2 \pi)^3 k_2^+}}  
\sqrt{2 (2 \pi)^3 P^+} \delta^3(P-k_1-k_2) \nonumber \\
&& ~~~~~\phi_2(P,\sigma \mid k_1, \sigma_1; k_2 , \lambda_2) b^\dagger(k_1,
\sigma_1) a^\dagger(k_2, \lambda_2) \mid 0 \rangle. 
\end{eqnarray} 
Explicit evaluation of Eq. (\ref{def1}) 
gives
\begin{eqnarray}
{F_2(x) \over x} && = \mid \psi_1 \mid^2 \delta(1-x) \nonumber \\
&& + \sum_{\sigma_1,\lambda_2} \int dx_2 \int d^2 \kappa_1^\perp \int d^2
\kappa_2^\perp \delta(1-x-x_2) \delta^2(\kappa_1^\perp + \kappa_2^\perp) 
\mid \psi_2^{\sigma_1, \lambda_2} (x,\kappa_1^\perp; x_2, \kappa_2^\perp)
\mid^2 \label{f2plus}
\end{eqnarray}
where we have introduced the Jacobi momenta ($x_i, \kappa_i^\perp$), $k_i^+ = x_i
P^+, k_i^\perp = \kappa_i^\perp + x_i P^\perp$, with $\sum_i x_i =1, \sum_i
\kappa_i^\perp =0$. We have also introduced the boost invariant amplitudes 
$\psi_1$ and $ \psi_2$  related to $\phi_1$ and $\phi_2$ by
$ \phi_1 = \psi_1$,
$ \phi_2(k_i^+, k_i^\perp) = { 1 \over \sqrt{P^+}} \psi_2 (x_i,
\kappa_i^\perp)$. The above equation makes manifest the parton
interpretation of the structure function $F_2$. 
To order $\alpha_s$ we have,
(for details see Ref. \cite{hk})
\begin{eqnarray}
{F_2(x) \over x} = \delta(1-x) + {\alpha_s \over 2 \pi} C_f ln{Q^2 \over
\mu^2} \Big [ {1+ x^2 \over 1-x} - \delta(1-x) \int dy {1 + y^2 \over 1-y}
\Big ]
\end{eqnarray}
which is the well-known answer.

Next we extract the structure function $F_2(x)$ from the transverse
component of the bilocal vector current. The constrained fermion field 
$ \psi^- = {1 \over i \partial^+} (\alpha^\perp.(i \partial^\perp + g
A^\perp) + \gamma^0 m) \psi^+$.
We use the light-front $\gamma$
representation given in Ref. \cite{wzhang}. 
Without loss of generality we take the $\perp$ direction along the $x$ axis.
The structure function can be explicitly written as
\begin{eqnarray}
{F_2(x) \over x} = { 1 \over 8 \pi} {P^+ \over P^1} \int dy^- 
e^{- {i \over 2} P^+ y^- x} \langle P \mid \xi^\dagger(y) \Big [
O_m + O_{k^{\perp}} + O_g \Big ] \xi(0) \mid P \rangle + h.c   ,
\end{eqnarray}
with
\begin{eqnarray}
O_m && = im { 1 \over i \partial^+} \sigma^2, \nonumber \\
O_{k^{\perp}} && = { 1 \over i \partial^+} \big [ i \partial^1 - \sigma^3
\partial^2 \big], \nonumber \\
O_g && = g { 1 \over i \partial^+} \big [ A^1 + i \sigma^3 A^2 ] \big .
\end{eqnarray}

Contribution from $O_m$: Only potential non-vanishing contributions are from
the diagonal matrix elements for the single quark state and the quark-gluon
state. Single quark matrix element vanishes because
$\sigma^2$ flips helicity. Diagonal contribution from the quark-gluon state
also vanishes because of the cancelation between the two terms in
Eq. (\ref{perp}).
Thus the contribution from $O_m$ to $F_2$ vanishes.

Contribution from $O_{k^\perp}$: Explicit evaluation leads to  
\begin{eqnarray}
{F_2(x) \over x}\mid_{k^\perp} && =\mid \psi_1 \mid^2 \delta
(1-x) \nonumber \\
&& + {1 \over P^1}  \sum_{\sigma_1,\lambda_2} \int dx_2 \int d^2 \kappa_1^\perp \int d^2
\kappa_2^\perp \delta(1-x-x_2) \delta^2(\kappa_1^\perp + \kappa_2^\perp) 
\nonumber \\
&& ~~~~~~~\mid \psi_2^{\sigma_1, \lambda_2} 
(x,\kappa_1^\perp; x_2, \kappa_2^\perp)\mid^2  { \kappa_1^1 + x P^1 \over x}
\nonumber \\
&& = \mid \psi_1 \mid^2 \delta
(1-x) \nonumber \\
&& +  \sum_{\sigma_1,\lambda_2} \int dx_2 \int d^2 \kappa_1^\perp \int d^2
\kappa_2^\perp \delta(1-x-x_2) \delta^2(\kappa_1^\perp + \kappa_2^\perp) 
\nonumber \\
&& ~~~~~~\mid \psi_2^{\sigma_1, \lambda_2} (x,\kappa_1^\perp; x_2, \kappa_2^\perp)
\mid^2  \label{kperp} 
\end{eqnarray}
since $
\int d^2 \kappa_1^\perp \kappa_1^1 \mid \psi_2 \mid^2 =0$
as a consequence of rotational invariance. Eq.(\ref{kperp}) gives
the same result as Eq. (\ref{f2plus}).

Lastly we evaluate the contribution from $O_g$:
\begin{eqnarray}
{F_2(x) \over x} \mid_g && = { 1 \over 2} 
{ g \over \sqrt{2 (2 \pi)^3 }} { 1 \over
P^1} \sum_{\sigma_1, \lambda_2} \int {dy \over \sqrt{1-y}} d^2
\kappa^\perp \chi^\dagger_\sigma \big [ \epsilon^1_{\lambda_2} + i \sigma^3
\epsilon^2_{\lambda_2} \big ] \chi_{\sigma_1} \nonumber \\
&& ~~~~~~~~\psi_2^{\sigma_1,\lambda_2}(y, \kappa^\perp; 1-y, -\kappa^\perp)
 + h.c. \nonumber \\
&& = 0.
\end{eqnarray}
This is because the quark-gluon amplitude $\psi_2$ has two types 
of  terms: a) terms proportional to the quark mass $m$ accompanied 
by $\sigma^\perp$ which vanish because $ \chi^\dagger_\sigma
\sigma^\perp \chi_\sigma = 0$, and  b) 
terms proportional to $\kappa^\perp$ which vanish because of rotational
symmetry. Thus the contribution from $O_g$ to the structure function
vanishes.

From Eq.(\ref{kperp}) and Eq. (\ref{f2plus}), it follows that 
the structure function extracted from Eq.(\ref{def2}) has the 
same result given by eq.(10) and hence the same parton 
interpretation as that extracted from Eq.(\ref{def1}). 
Thus we have explicitly demonstrated the parton interpretation of the
transverse component of the bilocal vector matrix element in unpolarized
deep inelastic scattering. The classification of twist in 
DIS or other hadronic collision processes based on the different 
components of light-front bilocal operators seems unreliable. 

The axial vector counterpart of the matrix element studied here enters the
transverse polarized structure function $g_T$. The presence of the Dirac
matrix $\gamma^5$ and the polarization vector $S^\mu$ leads to major 
differences. Recently, using the same calculational 
framework we have studied and clarified\cite{wh} the question 
of the physical interpretation of $g_T$. From Eq. (\ref{def0}) it
is clear that the hadron matrix element of the minus component of the 
bilocal vector current carries additional nontrivial information about the 
hadron structure and dynamics compared to the plus and transverse components.
It is of great interest to study this matrix element and its physical
implications for deep inelastic scattering.     


This work was performed while the authors were visiting the
International Institute of Theoretical and Applied Physics, Iowa State
University, Ames, Iowa, U.S.A. for an extended time.
This work was supported in part by the U.S. Department of Energy under Grant
No. DEFG02-87ER40371, Division of High Energy and Nuclear Physics.

\end{document}